\documentclass[review]{elsarticle}
\usepackage{lipsum}
\usepackage{graphicx}
\usepackage{epstopdf}
\usepackage{amsmath}
\usepackage{soul}
\usepackage{color}
\DeclareGraphicsExtensions{.eps}
\makeatletter
\def\ps@pprintTitle{%
 \let\@oddhead\@empty
 \let\@evenhead\@empty
 \def\@oddfoot{}%
 \let\@evenfoot\@oddfoot}
\makeatother

\begin{document}

\begin{frontmatter}

\title{On the role of Sm in solidification of Al-Sm metallic glasses}

\address[myfirstaddress]{University of Wisconsin-Madison, Department of Materials Science and Engineering,1509 University Ave, Madison 53706, USA}

\author[myfirstaddress]{G.B. Bokas}
\author[myfirstaddress]{L. Zhao}
\author[myfirstaddress]{J.H. Perepezko}
\author[myfirstaddress]{I. Szlufarska\corref{I. Szlufarska}}
\cortext[I. Szlufarska]{Author to whom correspondence should be addressed}
\ead{szlufarska@wisc.edu}

\begin{abstract}

During the solidification of Al-Sm metallic glasses the evolution of the supercooled liquid atomic structure has been identified with an increasing population of icosahedral-like clusters with increasing Sm concentration. These clusters exhibit slower kinetics compared to the remaining clusters in the liquid leading to enhanced amorphous phase stability and glass forming ability (GFA). Maximum icosahedral-ordering and atomic packing density have been found for the $\mathrm{Al_{90}Sm_{10}}$ and $\mathrm{Al_{85}Sm_{15}}$ alloys, respectively, whereas minimum cohesive energy has been found for the $\mathrm{Al_{93}Sm_{7}}$ which is consistent with the range of compositions (from $\mathrm{Al_{92}Sm_8}$ to $\mathrm{Al_{84}Sm_{16}}$) found experimentally with high GFA.

\end{abstract}

\end{frontmatter}


\par Metallic glasses (MGs) are known to have attractive mechanical properties such as high strength and high elastic limit, good corrosion and wear resistance and good biocompatibility~\cite{Inoue2000, Ashby2006, Perepezko2002c}. However, the low glass forming ability (GFA) of some MGs has so far hindered synthesis of large MG specimens such as Al-base systems and therefore limited engineering applications of these materials. Many efforts have been devoted towards developing alloys with high GFA~\cite{Inoue2000,Cheng2008b, Cohen1961}. In particular, a number of studies focused on understanding of the MGs atomic structure and its relation to the MGs' mechanical properties and GFA~\cite{Almyras2010, Bokas2013, Miracle2004}. It is well known that the MGs lack long range order but exhibit short- and medium-range orders~\cite{Miracle2004}. Commonly, short-range order (SRO) is characterized by Voronoi polyhedrons (VPs) and medium-range order (MRO) by the arrangement of VPs into networks~\cite{Cheng2009, Sheng2006, Lekka2012, Li2015a, Zhang2016}.

\par Among Al-based rare earth (Al-RE) metal alloys (RE=Y, La, Ce, Pr, Nd, Sm, Gd, Tb, Dy, Er or Yb), the Al-Sm binary alloy has been shown to exhibit GFA for the widest range of compositions (from 8at.$\%$ to 16at.$\%$ of Sm)~\cite{Inoue1998}. For this reason, many experimental studies have been performed on Al-Sm binary alloys prepared by rapid quenching~\cite{Wilde1999c, Kalay2010a, Perepezko2003, Perepezko2003a, Stratton2005, Wilde1999a}.  However, understanding the SRO structure and its evolution during solidification as well as understanding the effect of the Sm concentration on the GFA of Al-Sm binary MGs are still open questions. These questions are addressed by performing molecular dynamics (MD) simulations of Al-Sm MGs with a range of stoichiometries and by characterizing their atomic structure during the solidification process.


Simulations are performed using the Finnis-Sinclair empirical potential implemented for the Al-Sm MGs~\cite{Mendelev2015} in the LAMMPS software package~\cite{Plimpton1995}. Lattice parameters and formation energies of Al rich crystalline compounds have been included in the development of this potential. Although these parameters have been fitted only for Sm concentrations of up to 25at.$\%$, it will be shown that quantities, such as diffusion coefficient (DC), agree well with more accurate density functional theory (DFT) calculations for a wider range of compositions (up to 50at.$\%$Sm). Therefore $\mathrm{Al_xSm_{100-x}}$ alloys were analyzed with compositions in the range of $100 \le x \le 50$. The Finnis-Sinclair potential has the following functional form
$\mathrm{\mathit{U} = \sum\limits_{i=1}^{N-1}\sum\limits_{j=i+1}^N\theta_{t_it_j}(r_{ij})+\sum\limits_{i=1}^N\mathit{\Phi}_{t_i}(\rho_i)}$ where $\mathrm{t_i}$ represents species of atom $\mathrm{i}$, $\mathrm{N}$ is the total number of atoms, $\mathrm{r_{ij}}$ is the distance between atoms $\mathrm{i}$ and $\mathrm{j}$ and $\mathrm{\theta_{t_it_j}(r_{ij})}$ is the pairwise interaction energy. In the above equation, $\mathrm{\mathit{\Phi}_{t_i}(\rho_i)}$ is the embedding energy function and $\mathrm{\rho_i=\sum\limits_j\mathit{\Psi}_{t_it_j}(r_{ij})}$ is the electronic density, which in turn can be written as the sum of the electronic density functions $\mathrm{\mathit{\Psi}_{t_it_j}(r_{ij})}$ of the individual atoms. 
The starting system consists of 32,000 atoms arranged on a face-centered cubic (fcc) lattice with 20$\times$20$\times$20 unit cells. Periodic boundary conditions are applied in all three directions. Initially, Al and Sm atoms are randomly distributed on the fcc lattice so that the ratio of Al to Sm atoms is consistent with the selected alloy composition and then the system is equilibrated at 300K for 0.2ns. The sample is subsequently melted and equilibrated for 1ns at $T=2,000$K, several hundred degrees higher than the melting temperature. Simulations are performed in the isothermal-isobaric (NPT) ensemble where the temperature and pressure are controlled using the Nose-Hoover thermostat and barostat, respectively. Zero pressure is maintained throughout all simulations. Subsequently, each system was quenched to 2K in steps of 50K with three different cooling rates of $\mathrm{4 \frac{K}{ps}, 2 \frac{K}{ps}}$, and $\mathrm{1 \frac{K}{ps}}$. The properties of the systems have been calculated for all the cooling rates but most of the results (unless otherwise noted) will be presented for the cooling rate of $\mathrm{1 \frac{K}{ps}}$. At every 50K interval, NPT simulations are performed for 15ps in order to equilibrate the system. After this equilibration time, NPT simulations are continued at the same temperature for another 15ps in order to calculate average properties of the systems. The timestep in the simulations is 1fs. In addition, the calculated glass transition temperature $(T_\mathrm g$) and the melting temperature $(T_\mathrm{m})$ in the simulations were found to vary almost linearly from 520K to 1,125K and ${T_\mathrm{m}}$ monotonically from 933K to 1,400K, respectively, between 0 and 50at.$\%$Sm.

\par Smaller Al-Sm systems (256 atoms) were used for {\it ab initio} MD (AIMD) simulations based on DFT. These systems were prepared by classical MD simulations starting from an fcc lattice with 4$\times$4$\times$4 unit cells and periodic boundary conditions. Al and Sm atoms were distributed randomly on the fcc lattice. After that the systems were melted at $T=2,000$K and then cooled down to 2K, using classical MD simulations and NPT ensemble. Subsequently the samples were annealed inside the supercooled region by first gradually increasing the temperature to ${T=\mathrm{1.1 \times} T_\mathrm{g}}$ (below ${T_\mathrm{m}}$) and then annealing for 0.5ns. Samples prepared this way were then used as an input for AIMD simulations. DFT calculations were carried out using the Vienna Ab Initio Simulation Package (VASP)~\cite{Kresse1996, Kresse1999}. AIMD simulations were performed in a canonical ensemble (constant volume and constant temperature), where the temperature is controlled using the Nose-Hoover thermostat and the timestep for the AIMD simulations is 1.5fs. The $\Gamma$ point was used to sample the Brillouin zone of the supercell. The Projector Augmented Wave method and the Generalized Gradient Approximation~\cite{Perdew1993, Becke1988, Langreth1983} were used to describe the interacting valence electrons~\cite{Rajagopal1973, Perdew1996}. The purpose of these AIMD simulations is to calculate DCs and to demonstrate consistency for the empirical potentials. The samples were initially equilibrated by AIMD at ${T=\mathrm{1.1 \times} T_\mathrm{g}}$ for 55ps and then the DC was calculated from the mean square displacement during the subsequent 20ps of the simulations. Finally, the ground state energy of systems created by DFT at RT was calculated using the conjugate-gradient minimization method.

\par
\begin{figure}
\centering
\includegraphics[scale=0.4]{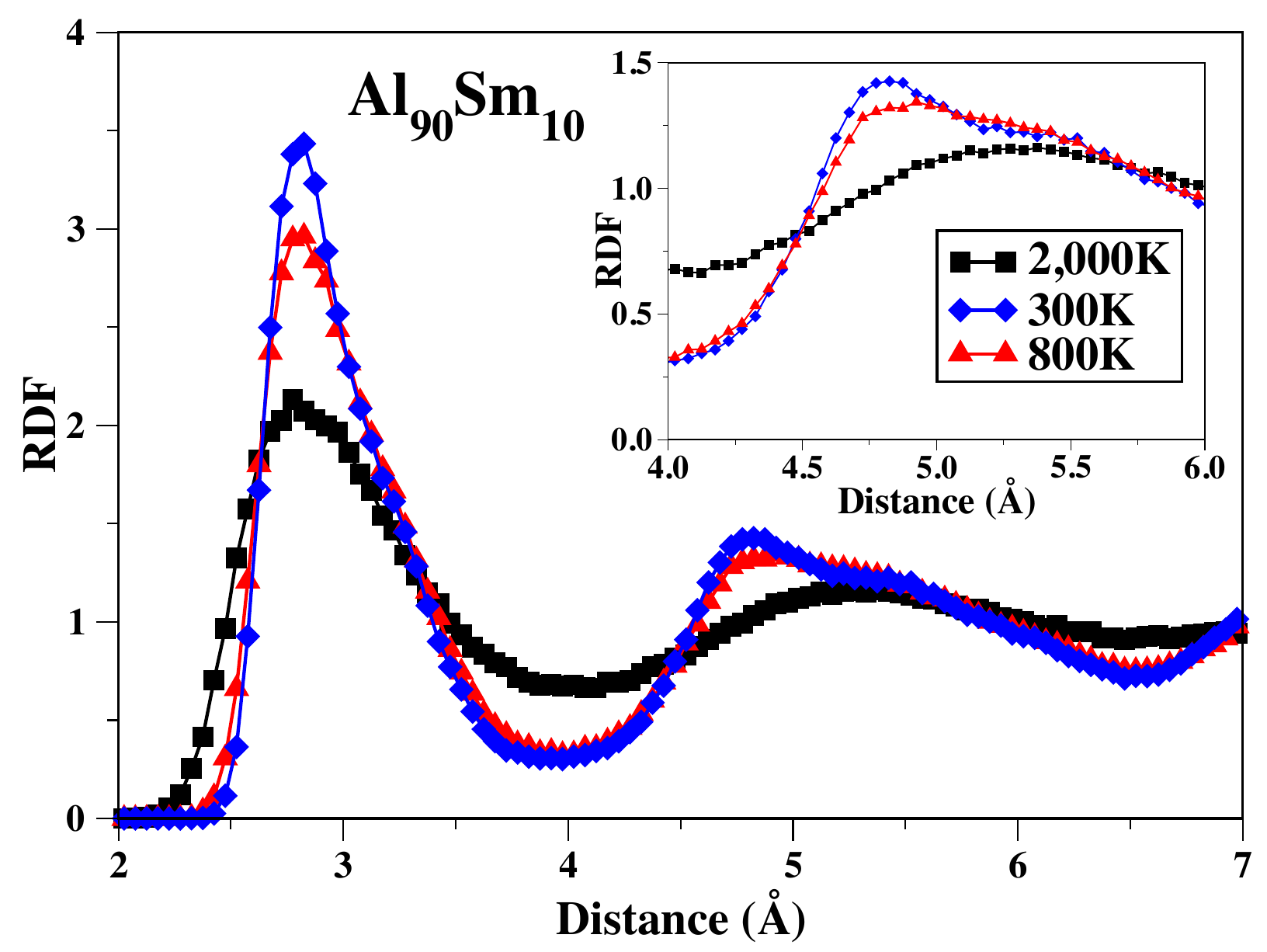}
\caption{(Color online) Total radial distribution function (RDF) of $\mathrm{Al_{90}Sm_{10}}$ for the liquid state at 2,000K, the supercooled region 800K, and the system after solidification at 300K. Inset: RDF of $\mathrm{Al_{90}Sm_{10}}$ showing the splitting at the second peak.}
\label{fig:rdf}
\end{figure}

 The formation of MG was verified by calculating evolution of the total radial distribution function (RDF) throughout the solidification process. An example of such evolution for $\mathrm{Al_{90}Sm_{10}}$ is shown in Fig.~\ref{fig:rdf}. Specifically, the RDF is given for the liquid state at 2,000K, supercooled system at 800K, and the system after solidification at 300K. The RDF is calculated at every $T$ by averaging the RDFs from the last 15ps of the NPT simulations. The plot in Fig.~\ref{fig:rdf} shows that the system at 2,000K is in a liquid state because the first peak is broad and there are no peaks at distances larger than 4$\mathrm{\AA}$. Upon quenching to 800K, the first nearest neighbor peak becomes higher and narrower, which suggests the initiation of some type of ordering. The second peak at this temperature around $\mathrm{5\AA}$ starts to split (see inset in Fig.~\ref{fig:rdf}). Such splitting is characteristic of MG structures that incorporate SRO and MRO~\cite{Group1977, Zhang2011, Lagogianni2009}. The sharpening of the first peak and splitting of the second peak are even more pronounced in the solidified structure quenched to room temperature (RT), which confirms MG formation. The RDF was also examined for other cooling rates studied in this work and found to have qualitatively similar distributions.

\par
\begin{figure}
\centering
\includegraphics[scale=0.4]{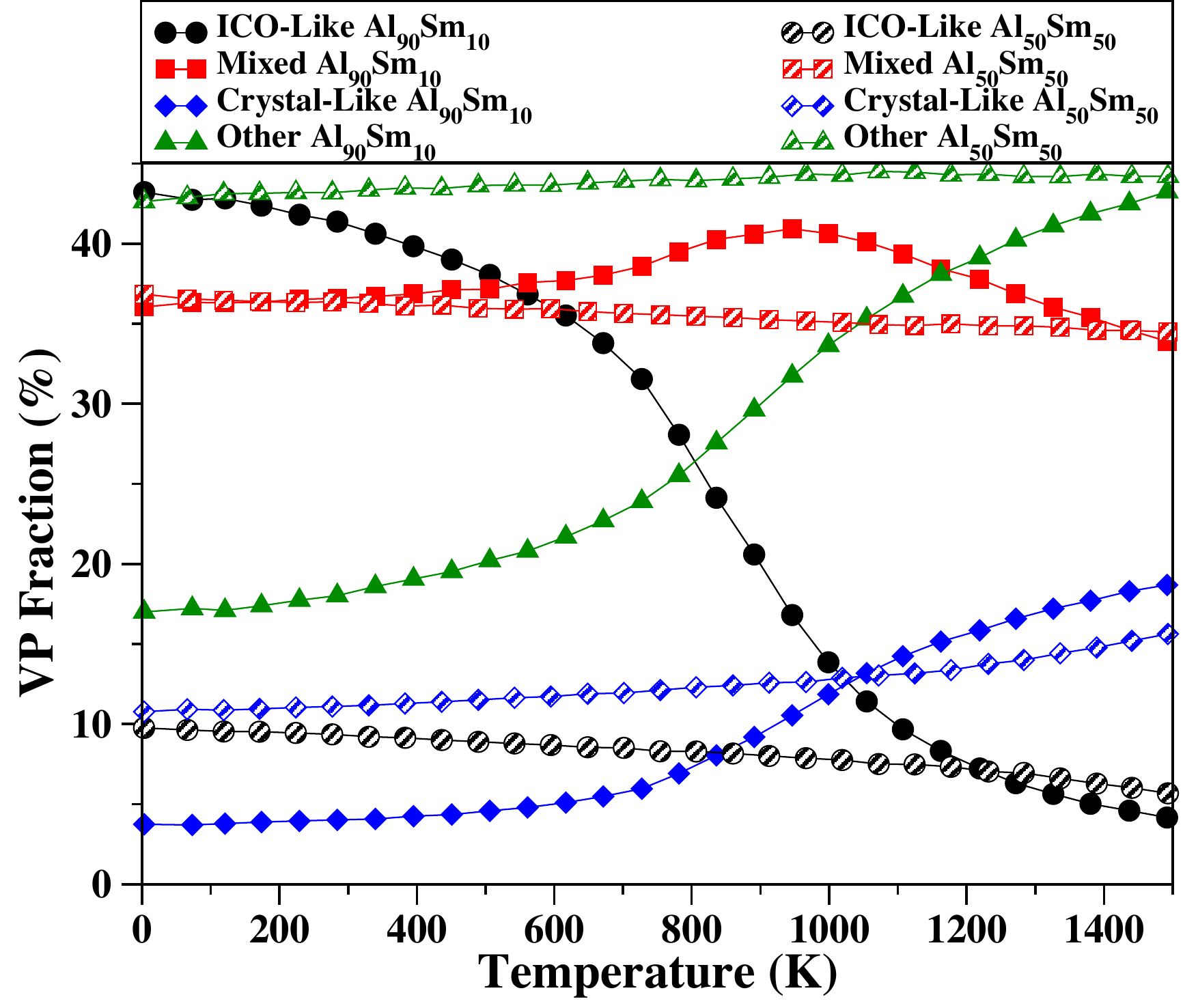}
\caption{(Color online) Evolution of ICO-like, crystal-like, mixed, and other VPs during cooling down for $\mathrm{Al_{90}Sm_{10}}$ and $\mathrm{Al_{50}Sm_{50}}$ (solid and half solid symbols, respectively).}
\label{fig:conc}
\end{figure}

 In order to gain further understanding into the effect of Sm on the atomic structure of the MG during solidification, the evolution of the atomic structure was monitored for various Sm concentrations as each system cools down to 2K. Voronoi tessellation technique~\cite{Finney1970} is a useful method to extract information about the system structure, especially the SRO. In this technique, a Voronoi polyhedral face is placed halfway between each pair of atoms. The edges of the polyhedron are defined by the intersection of these faces. The local structure can then be characterized by the number ($\mathrm{n_i}$) of the faces that contain a given number ($\mathrm{i}$) of the edges in each VP. Each particle of the system can be viewed as the center of a VP and thus the indices can be assigned at every system atom. The VPs are grouped as: icosahedral-like (ICO-like), crystal-like, mixed and other VPs. If the number of polyhedral faces with 5 edges is larger or equal to 10 ($n_5 \geq 10$) then the VPs are classified as ICO-like. If this number is smaller than 5 and at the same time the numbers of polyhedral faces with 6 and 4 edges is each larger or equal to 3, the VPs are labeled as crystal-like ($n_4 \geq 3$, $n_5 \leq 4$, $n_6 \geq 3$). Mixed VPs must have 3 polyhedral faces with 4 edges and 6 or more polyhedral faces with 5 edges ($n_4=3$, $n_5 \geq 6$). The remaining VPs are labeled as other VPs. The clusters exhibiting Al-like MRO are not treated in this analysis~\cite{Zhang2016, Stratton2005}. In Fig.~\ref{fig:conc} the fractions of the ICO-like, crystal-like, mixed and other VPs are shown versus temperature for $\mathrm{Al_{90}Sm_{10}}$ and $\mathrm{Al_{50}Sm_{50}}$. For the low Sm concentration systems ({\it i.e.}, $\mathrm{Al_{90}Sm_{10}}$), the fraction of mixed clusters changes only from 34$\%$ to 36$\%$ from 1,500K to 2K, and the fraction of the crystal-like clusters decreases from 18.5$\%$ to 3.5$\%$ in the same temperature regime. At the same time, the fraction of other VPs also decreases from 43.4$\%$ to 17$\%$. On the other hand, the fraction of the ICO-like VPs increases significantly from 4.1$\%$ to 43.5$\%$, especially inside the supercooled region, between $T_m\mathrm{=1,150K}$ and $T_g\mathrm{=700K}$. This increase of ICO-like VPs occurs mainly at the expense of other VPs, indicating the enhanced stability of the ICO-like clusters upon undercooling. Below $T_\mathrm{g}$ the fraction of the ICO-like VPs increases slower than above the $T_\mathrm{g}$ due to slower kinetics in the system.

Throughout the solidification process the population of the mixed VPs exhibits a broad and shallow maximum inside the supercooled region as illustrated in Fig.~\ref{fig:conc} (solid square symbols). As the temperature decreases below $T_\mathrm{m}$, the VPs form gradually into ICO-like clusters and therefore the number of pentagons inside the VPs increases. Between $T_\mathrm{m}$ and $\sim$900K, the population of VPs with 8 or more pentagons increases. According to the categorization, the VPs with $n_5 \geq 8$ belong to mixed and ICO-like groups and therefore both of these populations see an increase in numbers. Cooling the system further below $\sim$900K leads to an additional growth of the population of pentagons inside the VPs, which results in the growth of ICO-like clusters ($n_5 \geq 10$) at the expense of mixed clusters. In other words the VPs with $10\geq n_5 \geq 8$ can be viewed as a transitional population of VPs. From Fig.~\ref{fig:conc} it is also evident that ICO ordering has already started above the melting temperature of 1,150K. These findings imply that ICO-like VPs play a key role during the glass transition; in particular, inside the supercooled region. The evolution of the fraction of ICO-like clusters during the solidification procedure is in a qualitative agreement with findings in other alloys such as Cu-Zr~\cite{Zhang2011, Lagogianni2009, Mo2015, Duan2005, Cheng2008b}.

For the $\mathrm{Al_{50}Sm_{50}}$ alloy, the population of all the VPs remains approximately constant during quenching. For example, the populations of the ICO-like VPs and of the mixed VPs, respectively, change by less than 5$\%$ and 2$\%$ between 1,500K and 2K. The lack of increase of the number of clusters with fivefold symmetry inside the supercooled liquid of the $\mathrm{Al_{50}Sm_{50}}$ alloy is in contrast to the findings for the $\mathrm{Al_{90}Sm_{10}}$ alloy. This result suggests that the glass formation is not promoted in the $\mathrm{Al_{50}Sm_{50}}$ alloy. Finally, given that it was found experimentally that Al-Sm alloys with Sm concentration between 8at.$\%$ and 16at.$\%$ are good glass formers~\cite{Inoue1998}, these results further suggest that development of ICO-like clusters during quenching may be crucial for the vitrification of Al-Sm alloys.

\begin{figure}
\centering
\makebox[\textwidth][c]{\includegraphics[scale=0.5]{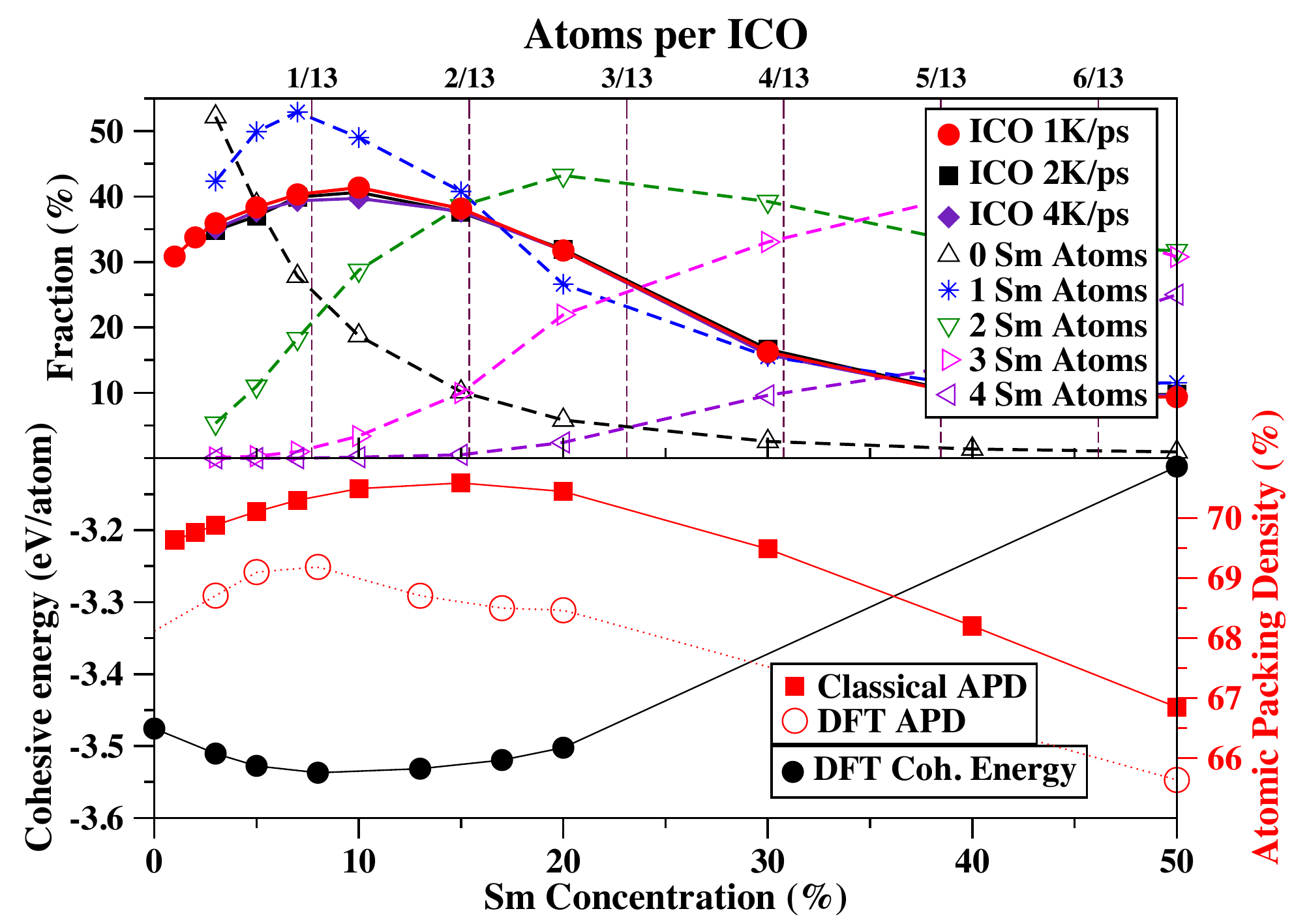}}
\caption{(Color online) (a) Solid symbols: fraction of ICO-like VPs for cooling rates $\mathrm{4 \frac{K}{ps}, 2 \frac{K}{ps}}$ and $\mathrm{1 \frac{K}{ps}}$, versus concentration of Sm (bottom horizontal axis). Open symbols correspond to the number of Sm atoms inside an ICO cluster, which consists of 13 atoms (top horizontal axis) at cooling rate $\mathrm{1 \frac{K}{ps}}$. The top and bottom horizontal axes are not independent of each other as explained in the main text. (b)  Cohesive energy calculation from DFT versus Sm concentration (black) and atomic packing density at 300K versus Sm concentration calculated by MD and DFT (red). The systems were quenched down with the cooling rate $\mathrm{1 \frac{K}{ps}}$.}
\label{fig:ICOs}
\end{figure}



\par To test this hypothesis, the fraction of ICO-like clusters at RT was calculated as a function of Sm composition. As illustrated in Fig.~\ref{fig:ICOs}(a), the highest icosahedral ordering is found at 10at.$\%$ of Sm regardless of the cooling rate used for vitrification of the alloy. Since the increased number of ICO-like clusters is often associated with better GFA, this finding is in good agreement with experimental results predicting the best GFA in Al-Sm alloys for concentration 8-16at.$\%$Sm~\cite{Inoue1998}. As the cooling rate decreases from 4K/ps to 1K/ps, there is a monotonic increase of the ICO-like population. Thus, icosahedral ordering is expected to be even more pronounced for the experimental MGs because their cooling rates are very low ($\sim$0.1K/ps~\cite{Inoue2000}), by comparison to the computational cooling rates.

Another interesting and related question is whether the number of Sm inside the Al-Sm ICO-like cluster depends on the composition of the entire sample. It has been argued~\cite{Almyras2010} that if there was a building unit ({\it e.g.}, ICO-like cluster, supercluster) of the MGs, its composition should be the same as the sample composition. This point is addressed by calculating the number of Sm atoms that exist inside the ICO-like clusters. In Fig.~\ref{fig:ICOs}(a), the top horizontal-axis represents the number of Sm atoms inside an ICO-like cluster (each such cluster consists of 13 atoms), while the bottom horizontal axis represents concentration of Sm atoms inside each cluster.
Dashed lines with open symbols correspond to the average number of Sm atoms that exist inside the ICO-like VPs for each composition of the sample. As the Sm concentration increases in the alloy, the number of Sm atoms inside the ICO-like clusters increases as well. There is only one Al-Sm composition ($\mathrm{Al_{93}Sm_{7}}$), where the Sm concentration inside most of the ICO-like clusters ($>50$at.$\%$, star symbols in Fig.~\ref{fig:ICOs}(a)) matches the $\mathrm{Al_{93}Sm_{7}}$ alloy composition.  At this particular composition, there is also a peak in the fraction of 1 Sm per ICO-like VP. The fact that the population of $\mathrm{Al_{12}Sm_1}$ ICO-like clusters has a maximum for the $\mathrm{Al_{93}Sm_{7}}$ system means that the compositions of the ICO-like VPs existing inside the $\mathrm{Al_{93}Sm_{7}}$ alloy have the tendency to become the same as the system's total composition. This result might be another indicator of enhanced glass stability for this particular alloy composition.

It is interesting to ask what other properties may correlate with compositions that have good GFA. It has been proposed that atomic packing density (APD) is one of them. APD determines the packing state of the VPs as well as the structural stability~\cite{Park2007} and it is considered to be an important factor that affects the MG's mechanical properties~\cite{Park2007}. APD is defined as the fraction of volume that is occupied by particles $\mathrm{APD=\frac{n_{Al}*V_{Al}+n_{Sm}*V_{Sm}}{Volume}}$ where $\mathrm{n_{Al}}$ and $\mathrm{n_{Sm}}$ represent the numbers of Al and Sm atoms, respectively, and $\mathrm{V_{Al}}$ and $\mathrm{V_{Sm}}$ are the atomic volumes of Al and Sm atoms. The APD of the systems at 300K calculated by MD was found to have a broad maximum at $\sim$15at.$\%$ Sm as shown in Fig.~\ref{fig:ICOs}(b) APD calculated by DFT after the conjugate-gradient relaxation exhibits a maximum at $\sim$8at.$\%$ of Sm. One should point out that APD is different from the average density of the alloy, which we found to be a monotonically decreasing function of Sm concentration. Another property that has been suggested to correlate with GFA is the cohesive energy~\cite{Laws2010},  which describes the strength of the bonds in the glass. We calculated cohesive energy using DFT and as shown in Fig.~\ref{fig:ICOs}(b) cohesive energy has a broad minimum at $\sim$10at.$\%$ Sm.

\par
\begin{figure}
\centering
\makebox[\textwidth][c]{\includegraphics[scale=0.6]{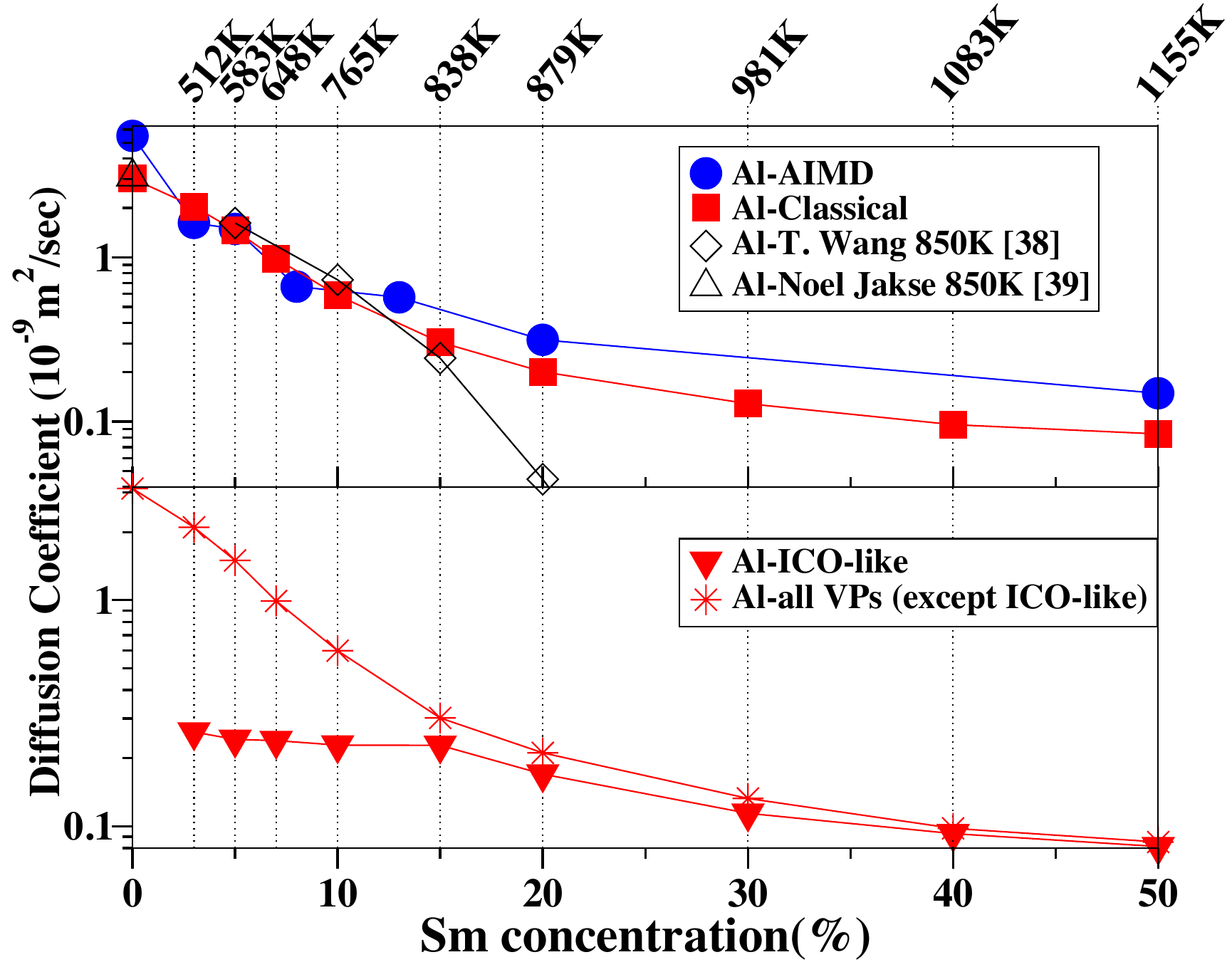}}
\caption{(Color online) (a) Diffusion coefficient calculated by classical MD and by {\it ab initio} MD for Al atoms versus Sm concentration. Top horizontal axis correspond to the temperature of the calculation, which is equal to 1.1 of Tg. Open symbols correspond to Al diffusion coefficient calculated from previous works at 850K.  (b) Diffusion coefficient calculated by classical MD for the ICO-Like VP Al central atoms (triangles) and for the remaining Al atoms in the sample (star symbols).}
\label{fig:prop}
\end{figure}

 In addition to structural properties of alloys, it has been proposed~\cite{Cheng2008b} that kinetic properties, such as the propensity for motion, of VPs might also affect the GFA of MGs. To test this hypothesis, the DCs of Al were calculated using both classical MD and AIMD calculations at $T=\mathrm{1.1 \times} T_\mathrm{g}$, i.e., inside the supercooled liquid. The results are shown in Fig.~\ref{fig:prop}(a) and they are compared with previous DCs calculated by AIMD at constant $T=850$K~\cite{Jakse2013, Wang2015}. DCs calculated by the two methods are in a good agreement in each other, which demonstrates that the empirical potential used here is reliable in reproducing kinetic properties of the Al-Sm alloys. The DCs of the Al atoms that lie at the center of ICO-like VPs are compared with the DCs of the remaining Al atoms in the sample. The central Al atoms have lower DCs than the DCs of the remaining Al atoms in the sample, especially for compositions below 15at.$\%$Sm (see inset in Fig.~\ref{fig:prop}). Thus, ICO-like VPs contribute to the stabilization of the amorphous phase due to their slow kinetics inside the supercooled region, for a specific Al-Sm composition. These results are in good agreement with previous studies~\cite{Cheng2008b}, which showed that the ICO clusters are responsible for the dynamical slowdown of the supercooled liquid. Finally, it is interesting to look at the trend in DC as a function of Sm concentration in the supercooled Al-Sm system. As shown in Fig.~\ref{fig:prop} there is a monotonic decrease of the diffusion coefficient with increasing Sm concentration. The monotonic change in diffusion coefficient is inconsistent with the experimental prediction that the best glass formers are in the range 8-16at.$\%$ of Sm and therefore this result indicates that the DC cannot be used as a glass former predictor across different alloy compositions.

In conclusion, during the solidification process, the Sm presence (until 10at.$\%$) was found to promote the icosahedral ordering in the supercooled region. At the same time the DCs of the central Al atoms of the ICO-like VPs is lower than the DCs of the remaining Al atoms in the sample. The increase of the ICO-like VPs population, due to the Sm presence, and the slower kinetics introduced by the ICO-like VPs in the alloys, results in slower relaxation dynamics inside the supercooled region. When ICO-like clusters are present, the alloys need more time to create crystal structure during solidification and thus these clusters can be correlated with better GFA and for the enhanced stability of the amorphous phase inside the supercooled region for Al-Sm alloys. Further addition of Sm (above 10at.$\%$) to the alloy does not lead to an additional increase of icosahedral ordering because the DC is very low. After quenching the alloy to RT, the Sm concentration inside the Al-Sm MG was found to have a significant effect on the icosahedral ordering, the atomic packing density and the cohesive energy. Specifically, there is a region between 10 and 15at.$\%$ Sm where the fraction of the ICO-like VPs and the atomic packing density are maximized, while the cohesive energy has a minimum in the same concentration regime. This outcome is consistent with experimental results that reported the glass formation range of the Al-Sm alloys as 8-16at.$\%$ Sm. Finally, the presence of structural motifs ({\it i.e.}, of ICO-like VPs) correlates strongly with GFA, whereas kinetic properties (such as diffusion) are not strong predictors of GFA.

\section*{Acknowledgments}

The authors acknowledge financial support from NSF-DMREF grant DMR-1332851.
 
\section*{References}

\bibliography{library}

\end{document}